\documentclass[aip,jcp,reprint]{revtex4-1}
\usepackage{graphics,graphicx,amsfonts,amsmath,amsbsy,amssymb,color}
\usepackage{mathrsfs}
\usepackage{multirow}
\usepackage{subdepth}
\usepackage{bm}
\usepackage{braket}
\usepackage{subfigure}
\usepackage{xspace}

\begin{document}
\author{Robert~E.~Thomas}
\email{ret41@cam.ac.uk}
\affiliation{University of Cambridge, The University Chemical Laboratory, Lensfield Road, Cambridge CB2 1EW, U.K.}
\author{Qiming Sun}
\affiliation{Department of Chemistry, Princeton University, Princeton, N.J. 08544, U.S.A.}
\author{Ali~Alavi}
\affiliation{University of Cambridge, The University Chemical Laboratory, Lensfield Road, Cambridge CB2 1EW, U.K.}
\affiliation{Max Planck Institute for Solid State Research, Heisenbergstra{\ss}e 1, 70569 Stuttgart, Germany}
\author{George~H.~Booth}
\email{george.booth@kcl.ac.uk}
\affiliation{King's College London, Department of Physics, Strand, London WC2R 2LS, U.K.}
\title{Stochastic multi-configurational self-consistent field theory}

\begin{abstract}

The multi-configurational self-consistent field theory is considered the standard starting point for
almost all multireference approaches required for strongly-correlated molecular problems.
The limitation of the approach is generally given by the number 
of strongly-correlated orbitals in the molecule, as its cost will grow exponentially with this number.
We present a new multi-configurational self-consistent field approach, wherein linear determinant coefficients of a multi-configurational wavefunction are 
optimized via the stochastic full configuration interaction quantum Monte Carlo technique at greatly reduced computational cost, 
with non-linear orbital rotation parameters updated variationally based on this sampled wavefunction. 
This extends this approach to strongly-correlated systems with far larger active spaces than 
it is possible to treat by conventional means.
By comparison with this traditional approach, we demonstrate that the introduction of stochastic noise in both the determinant 
amplitudes and the gradient and Hessian of
the orbital rotations does not preclude robust and reliable convergence of the orbital optimization. It can even improve the 
ability to avoid convergence to local minima in the orbital space, and therefore aid in finding variationally lower-energy solutions.
We consider the effect on the convergence of the orbitals as the number of walkers and the sampling time within the active space increases, as well as the effect
on the final energy and error.
The scope of the new protocol is demonstrated with a study of the increasingly strongly correlated electronic structure in a series of polycyclic aromatic 
hydrocarbons, up to the large coronene molecule in a complete active space of 24 $\pi$ electrons in 24 orbitals, requiring only modest computational resources.

\end{abstract}
   

\maketitle

\section{Introduction}

Much of the fields of computational electronic-structure theory and quantum chemistry is concerned with 
describing the changes in energies and properties due to the correlated motion of electrons.\cite{helgaker}
The treatment of these effects is, in general, required for the fields to be predictive and increasingly relevant 
to wider areas of chemistry, physics, and material science. This correlated physics of the electrons is generally
divided into two qualitative categories: {\em dynamic} and {\em static} correlation. 

As a broad categorization,
dynamic correlation describes the short-ranged, two-electron phenomena of the Coulomb holes and 
cusps denoting the depletion in probability amplitude of finding two electrons close to each other. Static correlation, on the other hand,
derives from the breakdown in the qualitative utility of the single-particle picture, which manifests as strong mixing between a number of determinants, 
with many-body co-operative effects giving rise to significant probability amplitudes of the wavefunction among its energetically low-lying configurations. If an 
electronic system contains significant static correlation, then perturbative approaches such as the ``gold standard'' of quantum chemistry --- coupled-cluster 
with singles, doubles and non-iterative triples (CCSD(T)) --- will necessarily fail, since they rely on the single-particle picture to generate
a single many-electron state about which to construct a perturbative expansion. This situation is prevalent in systems with localized, partially-occupied atomic 
states (such as the $d$-shells of transition metals), as well as low-spin, open-shell systems (including magnetic interactions) and important 
transition or stretched-bond configurations of molecular systems.

In quantum chemistry, it can be common to {\em define} the amount of static correlation in a problem as the difference between the uncorrelated Hartree--Fock energy
and that of complete active space self-consistent field (CASSCF) theory. The CASSCF method is a variational approach for optimizing a linear combination of low-energy
configurations which arise from a set of correlated ``active'' degrees of freedom, optimized self-consistently to provide the majority of the static portion of the correlation effects.\cite{roos1972,hinze1973,siegbahnheibergrooslevy1980,roostaylorsiegbahn1980,siegbahnalmlofheibergroos1981,knowleswerner1985,wernerknowles1985,olsen2011,LinDemkov2014}
This is then considered as the 
standard starting point for almost all multi-reference approaches required for strongly-correlated molecular problems.
A number of ``active'' orbitals and electrons are chosen, 
and the configurational subspace is constructed from all the possible distributions of those electrons in those orbitals, maintaining an inert ``core'' of the remaining particles.
The energy is then minimized within this configurational subspace. In addition to this, a unitary rotation of 
the non-redundant orbital space is found, again in a variational fashion, in order
to optimize the ``best'' orbitals, in a variational sense, to describe these configurations. This optimization can be crucial, since the original orbitals are generally based on the 
canonical Hartree--Fock orbitals, and hence do not necessarily span the qualitatively correct degrees of freedom to describe the strongly-correlated physics.

Since the high-energy part of the configurational spectrum is missing, this approach will not, with normal usage, be able to capture dynamic correlation from high-energy virtual scattering
effects to obtain a complete description of the correlated wavefunction. 
However, the final CASSCF wavefunction will achieve a qualitative description of a strongly-correlated molecular 
problem, which will have been previously missing if building upon a single-reference.
Dynamic correlation can often be included subsequently in the description on top of a CASSCF ``zeroth-order'' wavefunction in a perturbative or variational manner.\cite{anderssonMalmqvistroos1992,wernerknowles1988,shamasundar2011}

The greatest limitation upon the applicability of the CASSCF approach is the size of the active space which can be treated. While this space would, ideally, include all valence electrons and low-lying molecular orbitals,
the resultant number of variational parameters to optimize grows factorially with its size. 
Thus, computational bottlenecks usually demand that no more than around $16$ active electrons and degrees of freedom can be considered to contribute to the strong correlation.
For larger systems with multiple stronger correlation centres, this is often simply insufficient, and even for 
relatively small systems, testing the influence of additional effects such as `$s$-$p$' mixing or a double-$d$ shell to include $d$-$d$ transitions and 
wider energy scales in metal complexes is generally prohibitively expensive. 

In order to access these required larger space sizes, approximations are generally made to reduce the number of parameters which need to be optimized. The most prevalent of these are the
{\em restricted} active space,\cite{olsenroosjorgensenjensen1988,malmqvistrendellroos1990} and more recently the {\em generalized} active space constructions.\cite{fleig2001,fleig2003,ma2011,vogiatzis2015} 
These approaches choose only a subset of the complete active space configurations within which to optimize, thereby
dramatically cutting the number of configurations that need be considered. However, this adds additional uncertainties and approximations in the quality of the results, as well as additional 
complexity in setting up the calculation.

To maintain the full variational flexibility of the active space and the treatment of strong correlation effects which that entails, we adopt a different approach in this work. The optimization of the linear determinant coefficients is decoupled
from the non-linear orbital rotation parameters. The linear coefficients are then solved for stochastically each iteration using the machinery of the full configuration interaction quantum Monte Carlo (FCIQMC) method.\cite{booththomalavi2009}
Stochastically-derived quantities have been previously used for quantum-chemical methods,\cite{neuhauser2012,willow2013,neuhauser2014} 
and FCIQMC provides a probablistic approach to the solution of the configuration interaction (CI) problem, whereby sparsity in the wavefunction is exploited to allow for large numbers of determinants to be optimized in systems
for which a deterministic solution (or even enumeration of the space) would be impossible. Simultaneous to this sampling, one- and two-body reduced density matrices are accumulated, which are used in a 
subsequent step to form the gradient and Hessian for the orbital rotations amongst the active and inactive spaces. The rotations are then solved for
deterministically (using the stochastically sampled gradient and Hessian), since there are generally a manageable number of these variational parameters, which grow only as $\mathcal{O}\left[M^2\right]$ 
where $M$ is the total number of orbitals in the full space. This process is repeated until convergence, implicitly coupling the linear parameters of the active space back to the non-linear rotations within the full space.

We begin with a more detailed discussion of these steps, before comparing the new approach to traditional techniques. Of particular interest is how the systematic and random errors
of the result depend upon the number of walkers, 
and the upon amount of time for which these walkers stochastically sample the active space in the FCIQMC step, as well as an analysis of the nature of the convergence over updates of the non-linear
parameters. 
Moving beyond what can be feasibly optimized within traditional CASSCF, we turn to a series of increasingly large polycyclic aromatic hydrocarbons. These are important
systems in the development of new photovoltaic devices,\cite{Michl2013,Eaves2012} while the trend from molecular systems towards the infinite graphene system and its novel properties is also of interest.\cite{clar1964,havey1997}
However, as the conjugated $\pi$ system grows 
in size, there is no natural separation of energy scales from which an active space can be formed, and so the problem of a balanced choice of active space with which to compare 
systems of different sizes traditionally becomes difficult. 
With this new approach, however, we can consider the complete $\pi$ space in all cases, allowing for a consistent truncation of the correlation effects, and can examine trends 
as these systems grow in size. Density matrix renormalization group (DMRG) calculations have previously indicated a trend for increasing strong correlation effects, towards 
polyradical character, in one-dimensional polycyclic acenes and narrow nano-ribbons,\cite{hachmannchan2007,mizukami2013} and we compare this to the character of growth in 
two-dimensional polycyclic systems, up to the $\mathrm{C}_{24}\mathrm{H}_{12}$ coronene system, including all 24 $\pi$ electrons and orbitals in the active space.
It is found that, whilst there is a general increase in correlated behavior with increasing system size, the two-dimensional series also sees the development of degeneracies 
in the highest-occupied and lowest-unoccupied parts of the natural-orbital spectrum not observed in the linear cases.

\section{Methodology}

The wavefunction of MCSCF theory is ultimately expressed as a linear combination of $N$-electron Slater determinants, $\left\{\Ket{D_\mathbf{i}}\right\}$. Its crucial difference from a wavefunction expanded as a 
subset of configurations, however, is the inclusion of an orbital rotation operator, $\mathrm{e}^{-\hat{\kappa}}$, to allow for the iterative optimization of the space spanned by the correlated orbitals, in addition to 
the set of configurational coefficients, $\left\{C_\mathbf{i}\right\}$, within it:
\begin{equation}
\Ket{\Psi} = \mathrm{e}^{-\hat{\kappa}}\sum_\mathbf{i} C_\mathbf{i}\Ket{D_\mathbf{i}}.
\end{equation}
The anti-Hermitian operator, $\hat{\kappa}$, takes the form of a sum over antisymmetrized excitation operators, which in this work we take to be spin-free,
\begin{align}
\hat{\kappa}& = \sum_{pq} \kappa_{pq} \hat{E}^-_{pq}\\
& = \sum_{pq} \kappa_{pq}\left(\hat{E}_{pq}-\hat{E}_{qp}\right),
\end{align}
with
\begin{equation}
\hat{E}_{pq}=\hat{a}^\dagger_{p\alpha}\hat{a}_{q\alpha} + \hat{a}^\dagger_{p\beta}\hat{a}_{q\beta},
\end{equation}
such that the rotations are real and preserve spin. The task is thus to determine the coefficients, $\left\{C_\mathbf{i}\right\}$, and the matrix elements, $\left\{\kappa_{pq}\right\}$.
 
It is specifically the complete-active-space (CASSCF) treatment, which truncates the configurational space by partitioning the orbital space into inactive, active, and virtual subspaces, with which we are concerned here. No restrictions are placed on the allowed occupancies in the active manifold, but the inactive and virtual orbitals are constrained to be always doubly occupied and unoccupied, respectively, in all configurations. The related restricted-active-space (RASSCF) theory, which further divides the active space by placing restrictions on the number of holes and electrons allowed to appear at its lower and upper energetic limits, is also implemented, and will be considered in future work. In CASSCF, simplifications arise as all intra-space rotations (active-active rotations, for instance) are redundant, while all inter-space transformations are not.\cite{helgaker} The only transformations which need be considered, therefore, are inactive-virtual, inactive-active, and active-virtual rotations, which induces a sizeable reduction in the effective size of the matrix $\bm{\kappa}$. 

In principle, CASSCF may be undertaken as a simultaneous second-order optimization of the orbital and configurational parameters. Whilst some of the established CASSCF implementations take this approach,\cite{knowleswerner1985,wernerknowles1985,werner1987,shepard1987,schmidtgordon1998} 
it restricts the flexibility for using newly-developed solvers for strongly-correlated quantum problems to obtain the CI coefficients.
Increasingly, therefore, a two-step, macroiterative approach, in which the configurational and orbital parameters are optimized separately, has come to prominence.  
These approaches are predicated on the fact that the equations governing configurational optimization are equivalent to those found in ``stand-alone'' CI theory, and may thus be solved with any of the techniques which seek to approximate it. The approach using DMRG as the configurational solver has enjoyed considerable success,\cite{zgid2008,ghoshhachmannyanaichan2008,yanaikurashigeshoshchan2009,yanaichan2010,kurashigechanyanai2013} with studies on long-chain polyenes, $\beta$-carotene, and transition-metal clusters all proving fruitful, and the stochastic approach discussed here seeks, ultimately, to continue in that vein.

Our formulation uses the full configuration interaction quantum Monte Carlo (FCIQMC) technique as its configurational optimizer, which we have elucidated in previous work.\cite{booththomalavi2009,clelandboothalavi2010,boothalavi2010,boothclelandthomalavi2011,clelandboothoveryalavi2012,boothgrueneiskressealavi2013,thomasoveryboothalavi2014,boothsmartalavi2014,thomasboothalavi2015}
This approach achieves the stochastic integration of the underlying Schr\"{o}dinger equation \emph{via} an imaginary-time evolution of an ensemble of signed walkers in Slater-determinant space. 
This evolution is performed as an iterative application of ``spawning'', ``death'', and ``annihilation'' steps, with the effect that the walker populations on each determinant, $\left\{N_\mathbf{i}\right\}$, become proportional to the coefficient, $\left\{C_\mathbf{i}\right\}$, when averaged over the long-imaginary-time limit. To control the sign problem, the initiator approximation is used, which results in a method which exhibits a systematic error at too low
walker number, but which is systematically improvable to exactness as the walker number increases. Throughout this work, the initiator threshold parameter, $n_\mathrm{a}$, is chosen to be $3$.\cite{clelandboothalavi2010}
To further reduce the size of random error bars, we make use of the semi-stochastic adaptation, in which the projection operator for the ground-state is applied deterministically to a small, ``core'' space, and stochastically elsewhere.\cite{umrigar2012,bluntsmart2015} For the systems studied here, we choose our core spaces to be either all the determinants coupled to the reference, or $1000$ such objects, whichever is the smaller.

Crucially for our present purposes, the FCIQMC approach also facilitates the stochastically unbiased sampling of the two-body reduced density matrices (RDMs),\cite{overy2014,thomasopalka2015} given by
    \begin{align}
        \Gamma_{pqrs} &= \langle \Psi|\hat{a}_p^\dagger\hat{a}_q^\dagger\hat{a}_s\hat{a}_r|\Psi \rangle\\
                      &= \sum_{\mathbf{ij}}C_{\mathbf{i}}C_{\mathbf{j}}\langle D_{\mathbf{i}}|\hat{a}_p^\dagger\hat{a}_q^\dagger\hat{a}_s\hat{a}_r|D_{\mathbf{j}} \rangle, \label{eqn:2RDM}
    \end{align}
    from which the one-body reduced density matrix can be found as
        \begin{align}
        \gamma_{pq} &= \langle \Psi|\hat{a}_p^\dagger\hat{a}_q|\Psi \rangle     \label{eqn:1RDM}  \\
                    &=\frac{1}{N-1} \sum_a \Gamma_{paqa} .
    \end{align}
An unbiased sampling of these 	quantities demands the introduction of a second walker ensemble, to which the population dynamics are applied separately, and the statistics acquired independently, from the first. 
This ``replica'' sampling eliminates biasing by ensuring that all the required products of determinant amplitudes are calculated from populations from \emph{both} simulations, and has found application in a variety of related Monte Carlo techniques.\cite{zhangkalos1993,blunt2014,bluntalavibooth2015}

    The reduced density matrices allow for the calculation of a host of molecular properties including electrical moments, forces, and coupling to two-particle geminal functions,\cite{overy2014,thomasopalka2015,boothclelandalavitew2012} but their chief use in this 
    present study is in their access to the gradient, $\mathbf{g}$, and Hessian, $\mathbf{H}$, of the energy with respect to the orbital parameters,
    \begin{equation}
        g_{pq} = \frac{\partial E}{\partial\kappa_{pq}} = \Braket{\Psi | \left[\hat{E}^-_{pq},\hat{\mathcal{H}}\right]|\Psi},
    \end{equation}
    and
    \begin{align}
        H_{pq,rs} &= \frac{\partial^2 E}{\partial\kappa_{pq}\partial\kappa_{rs}}\\
                  &=\frac{1}{2}\left(1+\hat{P}_{pq,rs}\right) \Braket{\Psi|\left[\hat{E}^-_{pq},\left[\hat{E}^-_{rs},\hat{\mathcal{H}}\right]\right]|\Psi},
    \end{align}
    in which the operator $\hat{P}_{pq,rs}$ permutes the orbital pair indices $pq$ and $rs$, and $\hat{\mathcal{H}}$ is the relevant Hamiltonian.\cite{ghoshhachmannyanaichan2008,yanaikurashigeshoshchan2009,helgaker,LinDemkov2014} 
    These objects may be expressed entirely in terms of the reduced density matrices and the one- and two-electron integrals, $\left\{h_{pq}\right\}$ and $\left\{\left\langle pq || rs \right\rangle\right\}$, in much the same manner as the analogous quantities are formulated in Hartree--Fock theory. Indeed, convergence is defined when the gradient vanishes, at which point the generalized Brillouin theorem is satisfied, indicating that there is no interaction between the
CASSCF wavefunction and its singly-excited configurations.
    
The energies presented in this work are quasi-variational energy estimates, derived from these sampled density matrices, and given by
\begin{equation}
    E = \sum_{pq} h_{pq} \gamma_{pq} + \sum_{p>q, r>s} \Gamma_{pqrs} \left\langle pq || rs \right\rangle + h_{\mathrm{nuc}}. 
    \label{eqn:RDMEnergy}
\end{equation}
This ensures that consistency with the corresponding gradient and Hessian expressions is maintained. It would, however, be instructive in a future study also to consider two other, non-variational energy expressions commonly used
in FCIQMC, derived from a mixed estimator and the ``shift'' -- a measure of the offset required to yield a norm-conserving propagator.\cite{booththomalavi2009}

Although the matrix, $\bm{\kappa}$, may be obtained by solution of the stationary equation,
    \begin{equation}
        \mathbf{H}\bm{\kappa} = -\mathbf{g},
    \end{equation}
convergence is improved by instead constructing the \emph{augmented} Hessian and its corresponding eigenequation.\cite{lengsfield1980,lengsfield1982}. Thus, $\bm{\kappa}$ is found as the lowest root of the equation
\begin{equation}
\begin{pmatrix} \mathbf{0} & \mathbf{g}^\mathrm{T} \\ \mathbf{g} & \mathbf{H}/\lambda\end{pmatrix} \begin{pmatrix} 1/\lambda \\ \bm{\kappa}\end{pmatrix} = \bm{\epsilon}\begin{pmatrix} 1/\lambda \\ \bm{\kappa}\end{pmatrix}
\end{equation}
with $\bm{\epsilon} = \lambda\mathbf{g}^{\dagger}\bm{\kappa}$, whose solution is found at modest computational expense \emph{via} iterative subspace methods, as implemented in the {\tt PySCF} package.\cite{pyscf} 

The protocol outlined above represents a single macroiterative cycle -- a single FCIQMC convergence and update of the CASSCF energy and orbital coefficients. Once performed, the newly-obtained orbitals and transformed integrals are used to 
construct a new Hamiltonian, and the process continued until the desired convergence is achieved. The initial studies to which we now turn address the suitability of this approach for achieving meaningful and 
satisfactory convergences of this type, and exactly what level of convergence can be expected
of the macroiterations in the presence of random errors in the sampled energy, gradient and Hessian expressions.

\section{Results and discussion}

\subsection{Preliminary studies}

The natural first test for this new formulation is the comparison with entirely deterministic calculations. 
It ought, in principle, to provide the same results as an approach using conventional configuration-interaction theory for the configurational optimizer, 
and a comparison between the two thus yields some insight into the effects of the inevitable introduction of stochastic noise up the convergence. As concrete examples, 
we turn to the benzene and naphthalene molecules in their experimental equilibrium geometries and in cc-pVDZ basis sets.\cite{herzberg1966,dunning1989} The active spaces in 
these cases are (6,6) and (10,10), respectively, for which the diagonalization of the active Hamiltonian is straightforward. 

Of particular concern for the stochastic formulation is the satisfactory sampling of the density matrices. We return to a consideration of the behavior with number of walkers ($N_\mathrm{w}$) 
for the later systems, but our initial focus with these smaller molecules is to assess the effect of the number of iterations of the FCIQMC dynamic for the accurate accumulation of the
density matrices for these purposes, and the effect of this sampling time on the overall convergence. 

\begin{figure}[h]
    \begin{center}
        \includegraphics[width=0.5\textwidth]{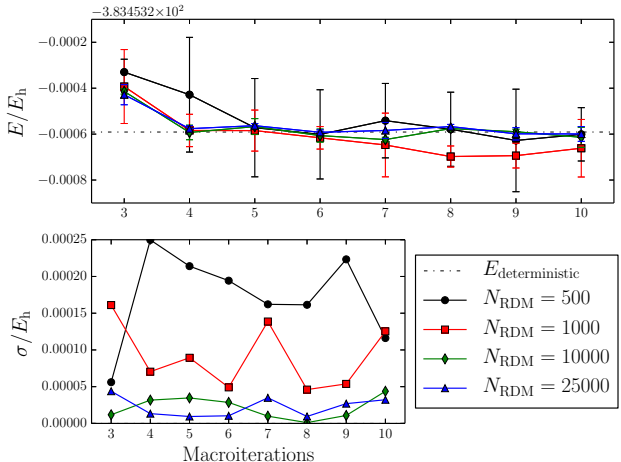}
    \end{center}
    \caption{Final stages of the macroiterative convergence for a (10,10) naphthalene calculation in a cc-pVDZ basis, with the configurational optimizations performed over different numbers of 
    density-matrix sampling iterations and using $10^4$ walkers. The error bars are obtained as the standard deviation, $\sigma$ of three independent simulations (not standard error), initialized with different random number seeds, and these are plotted in the lower panel. Too few iterations make 
    for noisy results, but the calculations performed with $25000$ sampling events are much more consistent, and converge to the deterministic result with satisfactorily small deviations once converged.}
\label{rdmiters}
\end{figure}

The convergence of the CASSCF macroiterations as the number of FCIQMC iterations, $N_\mathrm{RDM}$, for which the density matrices are sampled changes is shown in Figure \ref{rdmiters}. 
These RDMs are sampled afresh every macroiteration, and no prior initialization information is currently used between macroiteration 
steps, although the potential benefits of so doing are to be investigated in future studies.
It can be seen that the overall qualitative convergence profile of the macroiterations are relatively insensitive to the length of time sampling the density matrices. 
At each FCIQMC iteration, each walker has the possibility to sample only a single contribution to the sum in Equation \ref{eqn:2RDM} 
if a doubly excited walker is {\em successfully} spawned, and will contribute $N$ terms if a singly excited walker is created. 
It further contributes $N^2$ (diagonal) terms during the course of a configuration being occupied 
(this is accumulated and only included once during the lifetime of the determinant for computational expediency -- see Reference \cite{overy2014} for more details). 
The fact that as few as 500 sampling iterations with $10^4$ walkers (and a successful spawning rate of $\sim1\%$) can provide qualitatively correct results, therefore, demonstrates an efficient sampling for these purposes. This 
is substantially helped by the fact that the diagonal (generally dominant) part of the two-body RDM is averaged over
the stochastic wavefunction {\em exactly} for the sampling iterations, without requiring an additional stochastic step as is required for the off-diagonal elements in Equation \ref{eqn:2RDM}. 
Furthermore, any deterministic space defined as part of the `semi-stochastic' adaptation of the algorithm (which generally contains some of the largest weighted configurations) 
has its contribution to the density matrices included exactly.\cite{umrigar2012,bluntsmart2015,overy2014}
Whether this robust sampling is maintained or more iterations are required for systems with larger degrees of static correlation remains an open question, though this is likely offset by the larger number of walkers also required
to resolve the wavefunction.

Despite the similarity and robustness of the qualitative results, the number of sampling iterations does have the expected effect on the standard deviation in the energy estimator over 
independent samples at each macroiteration.
The variation between independent calculations, as measured by the standard deviations in Figure \ref{rdmiters}, can be traced to two effects due to the stochastic error in the density matrices. 
The first is the random error in the calculation of the variational energy for a given macroiteration as given by Equation \ref{eqn:RDMEnergy}, while the second is in the random error in the gradient and Hessian 
information in previous iterations, which gives different updates for the $\bm{\kappa}$ matrix and therefore orbital spaces in each macroiteration. 
Both of these effects are systematically controllable by changing
both $N_\mathrm{RDM}$ and $N_\mathrm{w}$.
Too few sampling iterations of the density matrices ($N_\mathrm{RDM}$) yields noisy behavior, and a standard deviation between independent calculations of up to $0.5\,mE_{\mathrm h}$. Increasing the sampling decreases the changes between energies at each macroiteration
as expected. The ultimate convergence criterion for the optimization of the overall wavefunction must be greater than this standard deviation, as even at ``convergence'', it is expected that the parameters will vary in a random
fashion of the order of this standard deviation. For this reason, the choice of $25000$ sampling iterations is chosen for the rest of the results in this work as providing close agreement with the fully deterministic result, and 
allows for a convergence threshold of the order $\pm 0.1\,mE_{\mathrm h}$.

\begin{figure}[h]
    \begin{center}
                \includegraphics[width=0.5\textwidth]{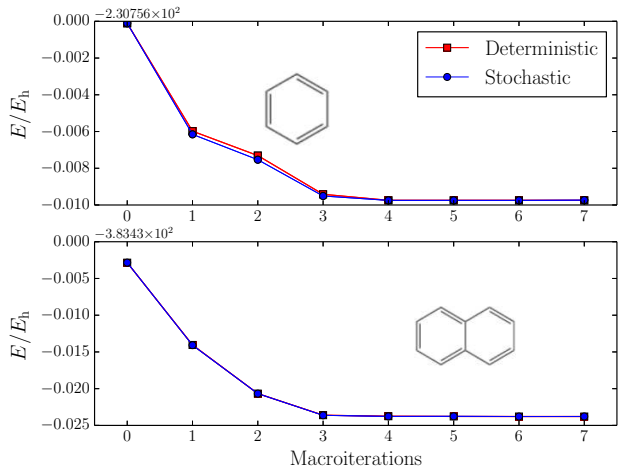}
                    \end{center}
                    \caption{The convergence of (6,6) and (10,10) calculations for benzene (upper panel) and naphthalene (lower panel) respectively, illustrating the close agreement between the deterministic and stochastic 
                    formulations. The energy of each stochastic macroiteration is given as the average of those of three independent calculations, and the error bars by the corresponding standard deviations. In both the (6,6) and 
                    the (10,10) examples, the convergence was achieved with $\mathcal{O}\left[10^5\right]$ walkers for each FCIQMC calculation. The zeroth iteration corresponds to the CASCI energy where the active orbital space is
                    made up of canonical Hartree--Fock orbitals, demonstrating the additional effect of the orbital optimization for these systems.}    
                    \label{benznaph}
                    \end{figure}

Having established that the density matrices can be obtained with sufficient quality for the orbital optimization step, we may compare the specific behavior of the stochastic convergence with that of its deterministic 
counterpart, as given in Figure \ref{benznaph}.  The conclusion borne out by these results appears to be that, once the density matrices are resolved with sufficient clarity, 
the stochastic noise inherent in this new formulation does not particularly change the convergence of the energy over that of the conventional approach. 
The stochastic method does, as is clear, demand a somewhat more permissive convergence criterion than its deterministic cousin --- of the order of $10^{-4}-10^{-5}\,E_\mathrm{h}$, as established above --- but the 
sub-millihartree agreement with deterministic results where comparison is possible suggests the feasibility of this new approach for addressing questions of physical and chemical significance. 

As an initial illustration of such problems, we consider the torsional barrier of the ethene molecule, C$_2$H$_4$, 
as an archetypal example of the type of substantial static correlation problem to which CASSCF is traditionally best suited.\cite{krylov1998} 
In this case, the bonding and anti-bonding $\pi$ and $\pi^*$ orbitals become degenerate as the dihedral angle, $\phi$, approaches $90^\circ$, 
and a proper treatment of the twisted geometry thus demands contributions from both the $\pi^2$ and the $\pi^{*2}$ configurations to give rise to a diradical singlet.

\begin{figure}[h]
    \begin{center}
                \includegraphics[width=0.5\textwidth]{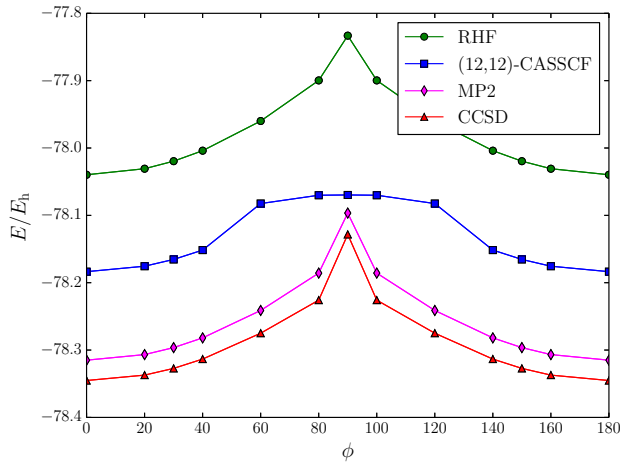}
                \end{center}
                \caption{The torsional barrier of the ethene molecule in a cc-pVDZ basis set at various levels of theory, with all other geometrical parameters held at their equilibrium experimental values.\cite{herzberg1966}.
                Cusp behavior indicates the inability to deal with the strong correlation effects which give rise to a diradical singlet character of the molecule at dihedral angles of $90^\circ$. The FCIQMC-CASSCF
            approach can consistently optimize the orbitals along this reaction coordinate in the presence of strong correlation effects, with $\mathcal{O}\left[10^6\right]$ walkers.}    
                \label{ethene}
            \end{figure}

Figure \ref{ethene} details the description of the torsional barrier by various levels of theory, and illustrates an important advantage of the CASSCF approach in its ability to eliminate the energetic cusp at $\phi=90^\circ$. This cusp appears in the restricted HF curve as a result of that theory's complete neglect of the excited configuration, and the inconsistent treatment of this state by both CCSD and MP2 fails to remove this feature completely. By contrast, the CASSCF allows for the necessarily equal contributions from the  $\pi^2$ and the $\pi^{*2}$ configurations at the barrier, and hence achieves the smooth curve required for a qualitatively correct description of the behavior. This also
demonstrates the ability for the FCIQMC-CASSCF to also work in situations with changing levels of correlation, robustly and consistently optimizing the orbitals across the potential energy curve. 

\subsection{Towards extended systems}

The ethene molecule provides a useful, albeit simple, illustration of the general importance of treating the full $\pi$-valence space when considering $\pi$-electron systems. The formulation introduced here, however, is applicable to active spaces of much greater size, and it is to these more extended systems which our attention now turns.

The polycyclic aromatic hydrocarbons, formed of fused six-membered rings,\cite{clar1964,havey1997} have long been a focus of both experimental and theoretical interest,\cite{angliker1982,kertesz1983,kivelson1983,wiberg1997,houk2001,bendikov2004a,bendikov2004b,bendikov2004c,mondal2006,reddy2006,hachmannchan2007}
providing technological potential,\cite{Michl2013,Eaves2012,reese2004} 
as well as being a source of medical and environmental concern.\cite{samanta2002,armstrong2004,srogi2007}
The linear acenes have been extensively studied with DMRG,\cite{hachmannchan2007,mizukami2013} as their quasi-one-dimensional structure is ideally suited to the method. 
However,
an advantage of FCIQMC is that its efficiency is not particularly dependent on the dimensionality of the system. As exemplar problems of this new technique, therefore, we explore a series of non-linear polycyclic aromatic hydrocarbons in cc-pVDZ basis sets,\cite{dunning1989} up to the coronene --- or superbenzene --- molecule, taking in the full carbon $\pi$ system in each case. This allows for a balanced active space for comparison between the systems
as they increase in size, without truncation of the conjugated electrons.
The geometries used are optimized at the level of density-functional theory,\cite{eckertpulaywerner1997,wernerknowles2012,molpro} using the B3LYP functional.\cite{vosko1980,leeyangparr1988,becke1993,stephens1994}

\begin{figure}[h]
    \begin{center}
                \includegraphics[width=0.5\textwidth]{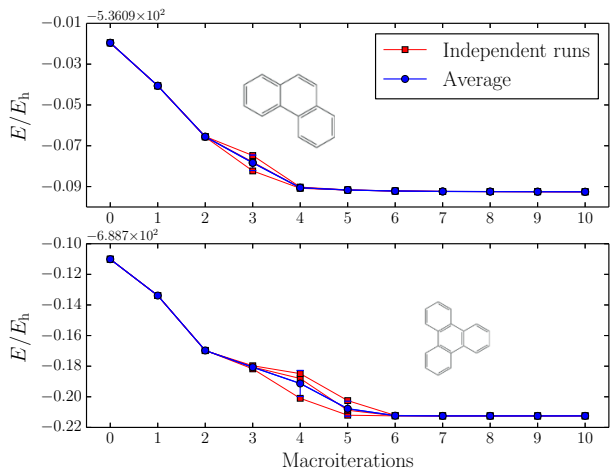}
                \end{center}
                \caption{Macroiterative convergences for phenanthrene (upper panel) and triphenylene (lower panel) in a cc-pVDZ basis.\cite{dunning1989} The phenanthrene system is studied with a (14,14) active space, in a total space of 246 orbitals, while for tripheylene we adopt an (18,18) active space in the 312-orbital total space. Each configuration iteration was allowed to evolve after equilibration for 25000 density-matrix sampling iterations. The curves shown in red detail the results of independent calculations, initialized with different random number seeds, while those shown in blue plot the averages of these simulations. While the independent convergences follow rather different trajectories, their latter iterations are in sufficient agreement with each other to give a small error bar for the obtained energy.}
                \label{phenanth-triphen}
            \end{figure}

Following on from the benzene and naphthalene studies of the previous section, we first consider the phenanthrene and triphenylene molecules, with (14,14) and (18,18) active spaces, respectively. 
This latter system is beyond the reach of deterministic CASSCF theories but, as illustrated in Figure \ref{phenanth-triphen}, the stochastic formulation provides a smooth, robust convergence.

The energy for each macroiteration is obtained directly from the reduced density matrices as in Equation \ref{eqn:RDMEnergy}, and, as in the previous section, we calculate an average energy for each point from the 
results of three independent calculations, initialized from different random number seeds, with the corresponding standard deviation serving as the error bar. 
As can be seen from Figure \ref{phenanth-triphen}, 
the precise trajectories of the independent convergences differ somewhat in their initial phases, as small random errors in the gradients and Hessians lead to propagation of differing orbitals through the macroiterative optimization.
However, the later iterations agree sufficiently well that a meaningful energy and error bar may be extracted, providing confidence that the same orbital space is sampled at convergence.
In particular, it is evident that performing each configurational step for $25000$ density-matrix sampling iterations after an equilibration period --- the rule of thumb inferred in the previous section --- 
is sufficient for the robust, reliable convergence which is our aim. In addition, walker populations were chosen such that the population residing on the reference exceeded $5\times 10^4$, which has previously been shown to be a 
satisfactory criterion for reliable convergence of any initiator error for systems with substantial weight on a single determinant.\cite{clelandboothoveryalavi2012,thomasoveryboothalavi2014,thomasboothalavi2015} This results in small changes to the total walker number for each iteration, 
as the wavefunction sparsity alters as the orbitals change. However, this was found to give consistently converged results for these systems, and we now turn to a more detailed consideration of the number of walkers 
required for larger active spaces in the coronene molecule.

\subsubsection{The effect of walker number}

It has been stressed that the macroiterative optimization of the orbitals in CASSCF is a non-linear problem, and as such
ensuring convergence to a global minimum is close to impossible, while sensitivity to initial conditions or simulation procedure can 
lead to differing solutions. One way in which convergence to global minima is encouraged in non-linear optimization problems is somewhat counterintuitively via the addition
of random noise into the gradient and Hessian in an optimization step. An example of this is the essential addition of ``noise'' to the 
optimization in DMRG, where a non-linear factorization of the wavefunction coefficients is optimized.\cite{white2005,zgid2008} 
Although the added noise in DMRG is not generally stochastic (though can be), neglect of this noise can lead 
to the convergence to meta-stable solutions and local minima, rather than the desired converged state.\cite{white2005} With this in mind, we 
detail the three anticipated effects of increasing the number of walkers in the FCIQMC-CASSCF procedure. 

First among these will be the decrease in ``initiator error'' in the FCIQMC
calculation.\cite{clelandboothalavi2010} The use of the initiator approximation (applied throughout this work) entails the introduction of a systematically improvable error into the
sampling of the wavefunction as a way of encouraging annihilation events and ameliorating the sign problem in the space. As the number of walkers increases, it can be
rigorously shown that the energy (generally rapidly) converges to the exact (FCI) result. Secondly, the increase in walkers will lead to a
reduction in the error in the sampled density matrices (and hence the gradient and Hessian of the orbital rotations), again tending to an exact limit. This is
due both to a reduction in the initiator error in the wavefunction from which the density matrices are sampled, and also an improvement in
the sampling of the matrices themselves. However, a third effect of changing the walker number relates to the convergence profile in the 
non-linear optimization of the orbitals. As the walker number increases, it will decrease the random noise in the sampled gradient and
Hessian information for the orbital rotation steps, and this will influence the stability of the global optimization of the orbitals.

\begin{figure}[h]
    \begin{center}
                \includegraphics[width=0.5\textwidth]{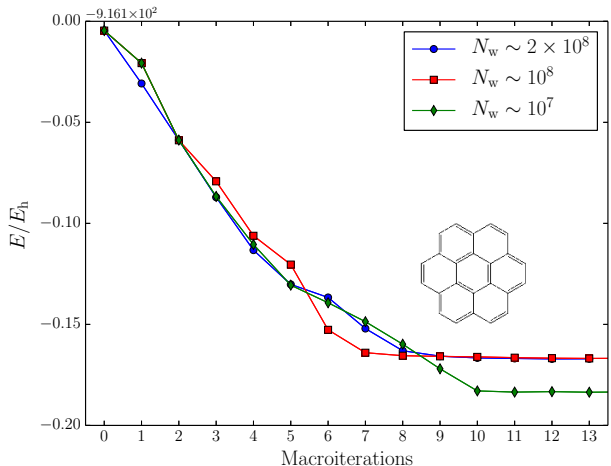}
                \end{center}
                \caption{Convergences for the coronene molecule in a cc-pVDZ basis, using different walker populations, $N_\mathrm{w}$, for each configurational optimization, in a (24,24) active space chosen from a $396$-orbital total space. Walker populations over $50$ million result in robust convergence to a meta-stable state, but an undersampled calculation using $N_\mathrm{w}\sim 10^7$ yields a lower-energy orbital solution. This solution can then be refined by increasing the walker number (see Fig.~\ref{restart}).}
                \label{coronene}
            \end{figure}

The convergences for coronene, using a (24,24) active space and different values for $N_\mathrm{w}$, are shown in Figure \ref{coronene}. The number of active space determinants is $\mathcal{O}\left[10^{12}\right]$. 
The simulations which are carried out with $100$ and $200$ million walkers per FCIQMC calculation (the latter value of which fulfils the criterion of placing $5\times 10^4$ walkers upon the reference) 
provide the kind of smooth, robust convergences in close agreement with one another which we observe for 
the smaller systems. Although slightly different optimization pathways may be taken, these walker numbers both achieve a quantitative agreement at convergence within a small standard deviation of each other.
However, a calculation using only $10$ million walkers for each step, results in a very different convergence to a lower-energy solution, in marked disagreement with the other results.

\begin{figure}[h]
    \begin{center}
                \includegraphics[width=0.5\textwidth]{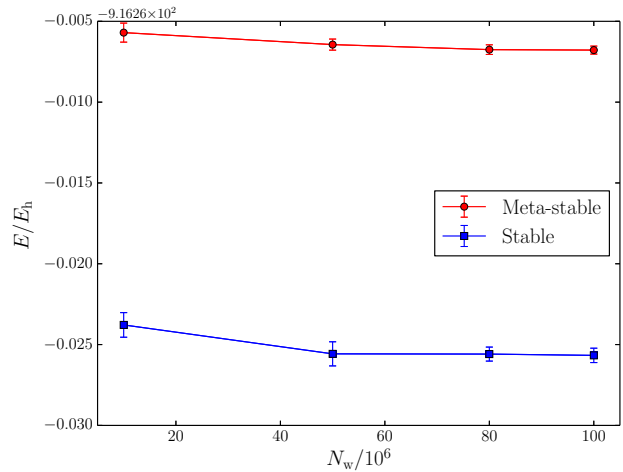}
                \end{center}
                \caption{FCIQMC energies for the meta-stable (red) and stable (blue) states found for coronene, using the converged orbitals from separate CASSCF optimizations with different walker populations, $N_\mathrm{w}$. The meta-stable orbitals correspond to the optimization at 200 million walkers, while the stable state corresponds to 10 million walkers, as shown in Fig.~\ref{coronene}. The removal of initiator error as $N_\mathrm{w}$ is increased decreases the energy of both states, but is not sufficient to explain the discrepancy between the two.}
                \label{init-error}            
\end{figure}

This discrepancy is not merely a simple manifestation of initiator error, but rather a clear example of the sensitivity of the optimization to its initial and early conditions.
As Figure \ref{init-error} illustrates, increasing the number of walkers in FCIQMC calculations to eliminate any effects of initiator error using the converged orbitals of both the meta-stable (higher energy) and 
stable (lower energy) states decreases their energy as the initiator error is removed, but the two states remain well separated by $\sim20mE_{\mathrm{h}}$, while the initiator error is less than $2mE_{\mathrm{h}}$.

\begin{figure}[h]
    \begin{center}
                \includegraphics[width=0.5\textwidth]{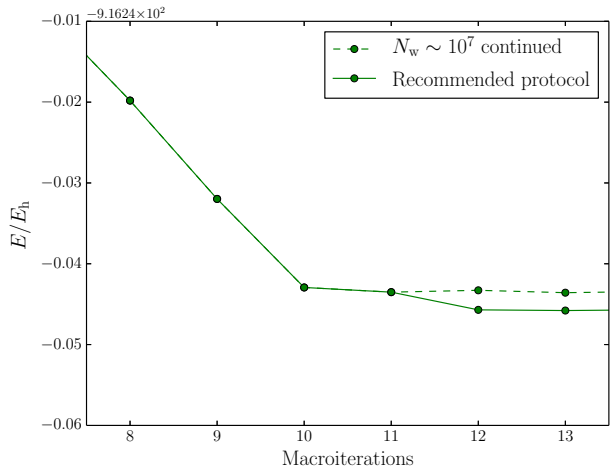}
                \end{center}
                \caption{Convergence for the (24,24) coronene system with the initial macroiterations performed using $10^7$ walkers, and those after and including macroiteration $12$ using $10^8$, yielding a result in agreement with the analysis of Figure \ref{init-error}. The dashed curve illustrates the trajectory if a walker population of $\mathcal{O}\left[10^7\right]$ is maintained in these latter stages, giving an indication of the initiator error encountered upon so doing.}
                \label{restart}    
\end{figure}

It appears, therefore, that rather noisy initial macroiterations may, in fact, be advantageous in converging upon a true ground state, much in the manner that added noise is helpful for DMRG. 
Indeed, in Figure \ref{restart}, we adopt a protocol wherein the first configurational optimizations are performed with $\mathcal{O}\left[10^7\right]$ walkers,
and macroiteration $12$ and later steps with $100$ million walkers to converge residual initiator error. 
This approach yields a trajectory towards the converged energy of the stable state in Figure \ref{init-error},
and is the recommended procedure for future studies to encourage convergence to the global minima of the orbital optimization problem. We speculate further that a deterministic FCI solver of the (24,24) coronene problem 
(if possible) with the same CASSCF algorithm would also likely converge to the higher energy orbital solution, and that for large active spaces or in the presence of strong correlation and difficult convergence, the addition of
artificial noise to the CASSCF optimization may also yield benefits in convergence stability. The protocol of optimizing first at lower walker numbers also reduces the computational cost of an FCIQMC-CASSCF calculation, since
only a few final iterations are required at the higher walker number finally to eliminate initiator error. While the optimisation is weakly exponentially scaling with the number of orbitals and electrons in the active space,\cite{boothclelandthomalavi2011,clelandboothoveryalavi2012,boothsmartalavi2014,shepherdscuseriaspencer2014} in the example of the (24,24) coronene system 
in Figure \ref{restart}, this corresponded to only $12$ hours on $320$ cores per macroiteration (including savings due to symmetry), taking advantage of the highly parallelizable nature of the FCIQMC algorithm. Memory requirements were also small, at $\sim3$GB of distributed memory (in contrast to the unfeasible traditional calculation which would require a minimum of $\sim100$TB of RAM). 

\subsubsection{The electronic structure of polycyclic hydrocarbons from FCIQMC-CASSCF}

Further insight into the electronic structure of these systems may be gleaned by considering the natural orbital occupation numbers, $\left\{n_i\right\}$, of the converged wavefunctions. 
Integrated measures of the number of ``effectively unpaired'' electrons in a system have been proposed by Takatsuka,\cite{takatsukafuenoyamaguchi1978,takatsukafueno1978,bochicchio1998,staroverov2000a,staroverov2000b} as
\begin{equation}
    N_{\mathrm{unpaired}}^{\left(\mathrm{T}\right)} = \sum_i \left(2n_i - n_i^2\right),
\end{equation}
and by Head-Gordon,\cite{headgordon2003a,headgordon2003b,bochicchio2003} as
\begin{equation}
N_{\mathrm{unpaired}}^{\left(\mathrm{HG}\right)} = \sum_i \min\left(n_i,2-n_i\right),
\end{equation}
in which $0\leq n_i \leq 2$. In addition, the HONO-LUNO gap, $n_\mathrm{HONO}-n_\mathrm{LUNO}$, can be used, where
HONO and LUNO correspond to the highest/lowest occupied/unoccupied natural orbital respectively. This quantity should not be considered as a measure of an energetic excitation, but rather as a metric for the deviation 
from the suitability of a Hartree--Fock picture of the electronic wavefunction.
As has been previously pointed out,\cite{hachmannchan2007,mizukami2013} 
a little care must also be exercised in interpreting the values of $N_\mathrm{unpaired}$; certainly, they should not be taken literally as the number of unpaired electrons found in a given system, but, taken together with $n_\mathrm{HONO}-n_\mathrm{LUNO}$, provide a useful indication of the correlated effects present in differently-sized molecules.

\begin{figure}[h]
    \begin{center}
                \includegraphics[width=0.5\textwidth]{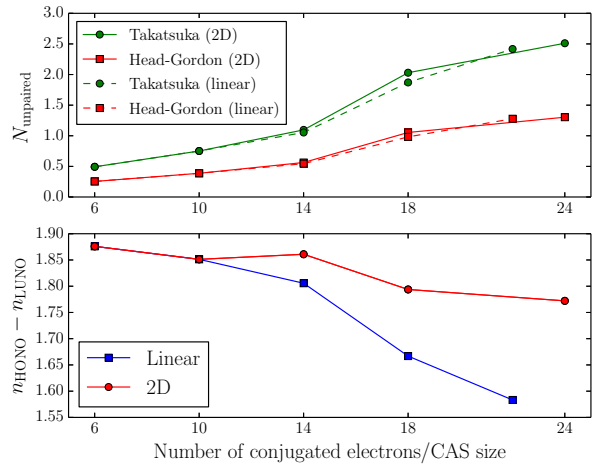}
                \end{center}
                \caption{Integrated metrics $N_{\mathrm{unpaired}}^{\left(\mathrm{HG}\right)}$ and $N_{\mathrm{unpaired}}^{\left(\mathrm{T}\right)}$, along with the HONO-LUNO gaps, for the two-dimensional systems considered here, compared with the series of linear acenes of similar size.}
		\label{unpaired-gap}
            \end{figure}

Figure \ref{unpaired-gap} compares these metrics for the two-dimensional systems discussed thus far to the corresponding linear acenes, 
which we have also calculated. The trend for the linear systems reveals, as has been previously reported,\cite{hachmannchan2007} 
the steady increase in correlated behavior as the chain length (and CAS size) is increased, leading to increasingly polyradical character of the wavefunction. 
The behavior for the two-dimensional systems, however, is rather less straightforward, indicating the strength of the correlation effects and radical nature of the resulting wavefunction is not simply 
related to the number of $\pi$ electrons in the conjugated system, but also to the precise geometry of the molecule under study. Whilst the integrated measure of the radical character is approximately the same for the
1D and 2D structures when comparing the same number of conjugated electrons, the HONO-LUNO gap does not increase at the same rate.

\begin{figure}[h]
    \begin{center}
                \includegraphics[width=0.5\textwidth]{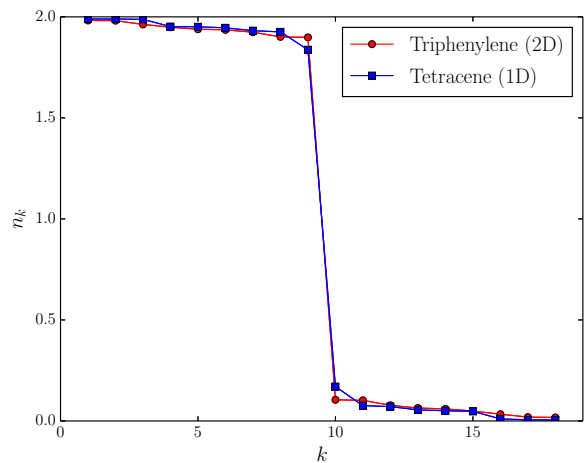}
                \end{center}
                \caption{Natural orbital occupation numbers for the (18,18) active spaces of tetracene and triphenylene, illustrating the degeneracy of the HONO and LUNO in the two-dimensional case, which is not present in the linear system.}
                \label{nos-comparison}            \end{figure}

\begin{figure}[h]
    \begin{center}
        \includegraphics[width=0.5\textwidth]{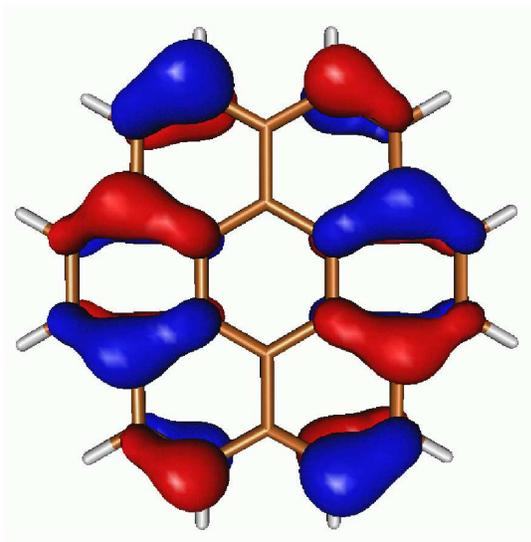}
    \end{center}
    \caption{Visualization of the highest-occupied natural orbital for coronene\cite{molden}, illustrating the accumulation of electron density around the outer carbon sites.}
    \label{hono}
\end{figure}

This behavior can be rationalized by examining the occupation-number spectra, as shown in Figure \ref{nos-comparison} for linear tetracene and two-dimensional triphenylene, which have the same number of electrons
in its $\pi$-system. These exemplar spectra highlight that, whilst the HONO-LUNO gap of the linear system is smaller than that of the two-dimensional case, 
the latter system sees the emergence of degeneracy in both the HONO and the LUNO not present in the former, contributing to the integrated measures of the radical nature. This is expected due to the increased
symmetry of the 2D system, as compared to the linear acene. It has previously been shown that the radical HONO and LUNO orbitals are related to {\em edge} effects of these conjugated systems, with
larger systems exhibiting HONO and LUNO orbitals in these systems which increasingly localize onto the perimeter of conjugated system.\cite{mizukami2013} This rationalizes the results of Figure \ref{unpaired-gap}, where
the smaller perimeter of the two-dimensional systems compared to the linear acenes for the same number of $\pi$ electrons is reflected in the size of the HONO-LUNO gap. Visualization of the coronene HONO in Figure \ref{hono}, also indicates that these orbitals that are most radical in nature are indeed beginning to localize on the edge of the molecule. It is anticipated that the increased degeneracy
of the 2D conjugated systems will mean that for a given {\em perimeter} of these systems, the 2D systems will exhibit stronger correlation effects than their 1D counterparts. Theoretically observing these
correlated trends provides a stern challenge for future applications of this new approach. However, the applicability and relative ease in dealing with these large 
correlated spaces with a stochastic solver bodes well for its future utility and scope to probe these systems.

\section{Conclusions}

A novel formulation of multi-configurational self consistent field theory, 
using a two-step, macroiterative algorithm with full configuration interaction quantum Monte Carlo for its configurational optimizations,
is capable of performing successful calculations with large active spaces. In this work, this ability is demonstrated with a study of a series of two-dimensional polycyclic aromatic 
hydrocarbons, culminating in calculations on the coronene molecule with a (24,24) active space. 

The inevitable introduction of stochastic noise in this approach does not significantly worsen convergence over a deterministic method, provided that satisfactory sampling is undertaken. In particular, we observe that the FCIQMC calculations must be performed for sufficiently many density-matrix sampling iterations (with $25000$ a useful, if informal, lower bound) and with sufficiently many walkers if the adverse effects of errors in the 
density matrices are to be avoided. 
Moreover, it is shown that the noise brought about by using comparatively few walkers for the initial iterations can be of benefit in converging to a true ground state, with high populations only required in the 
last stages of convergence in order to provide a satisfactory treatment of initiator error. 
The modest computational resources thus required suggest that
larger active spaces still will be within range, and thus that major new avenues of inquiry --- including studies of transition-metal compounds and clusters --- are now open and navigable. Furthermore, an approach for obtaining
excited states within FCIQMC will allow for a {\em state-averaged}-CASSCF implementation in the future, allowing for studies of excitations and photochemistry in correlated molecules.\cite{bluntsmartboothalavi2015}

\section{Acknowledgements}

The authors wish to thank Nicholas Blunt for helpful comments on the manuscript.
R.E.T gratefully acknowledges Trinity College, Cambridge for funding, while G.H.B gratefully acknowledges the Royal Society via a University Research Fellowship.  This work was supported by EPSRC under Grant No. EP/J003867/1. The calculations made use of the facilities of the Max Planck Society's Rechenzentrum Garching.

\providecommand{\latin}[1]{#1}
\providecommand*\mcitethebibliography{\thebibliography}
\csname @ifundefined\endcsname{endmcitethebibliography}
  {\let\endmcitethebibliography\endthebibliography}{}

\end{document}